\documentclass[matbio, referee]{svjour}

\usepackage{graphicx}
\usepackage{amsmath}

\title{SIR dynamics in structured populations with heterogeneous connectivity}
\titlerunning{SIR dyn. in struct. pop. w. heter. conn.}
\author{Erik Volz}

\institute{Department of Sociology, Cornell University, Ithaca, NY 14850,~\email{emv7@cornell.edu}}

\date{Received: September 4, 2005, Revised:\today}

\begin{document}
\maketitle

\abstract{
Most epidemic models assume equal mixing among members of a population. An alternative approach is to model a population as a random network in which individuals may have heterogeneous connectivity. This paper builds on previous research by describing the exact dynamical behavior of epidemics as they occur in random networks. 
A system of nonlinear differential equations is presented which describes the behavior of epidemics spreading through random networks with arbitrary degree distributions. The degree distribution is observed to have significant impact on both the final size and time scale of epidemics.
}

\section{Introduction\label{sec:intr}}

Contact patterns constitute an important aspect of heterogeneity within a population of susceptible and infectious individuals. It has also been one of the hardest factors to incorporate into epidemiological models. Compartment models have been able to capture many aspects of population heterogeneity, such as with respect to heterogeneous susceptibility and infectiousness~(\cite{veli1,and,die}). But compartment models can be inadequate with respect to population structure, especially when contact rates follow a steep and continuous gradient. 

Network theory describes a population of susceptible and infectious individuals as nodes in a network~(\cite{lilj1,strog1,newm2,andeMay2}). This has spawned a new category of epidemiological models in which epidemics spread from node to node by traversing network connections~(\cite{satoVesp1,meyePourNewmSkowBrun1,newm1,warr1,bara1,saraKask1}). Pure random networks with specified degree distributions have been proposed as realistic models of population structure. This case has the advantage of being well understood mathematically. The limiting behavior of epidemics spreading through random networks with a given degree distribution has been solved exactly~(\cite{meyePourNewmSkowBrun1,newm1}). 

The network approach has the advantage that the mathematics of stochastic branching processes~(\cite{wilf1,harr1,athrNey1}) can be brought to bear on the problem. This allows for precise descriptions of the distribution of outbreak sizes early in the course of the epidemic, as well as a solution for the final size of epidemics~(\cite{meyePourNewmSkowBrun1,newm1}).

A shortcoming of the network model has been that stochastic branching processes are inadequate to describe the explicit dynamical behavior epidemics. Thus the distribution of outbreak sizes are easy to solve for, yet the incidence curve, that is the number of infecteds at a time t, has been difficult to derive. Simulation has been used in this case~(\cite{euba1}). 

Heterogeneous networks make it difficult to derive differential equations to describe the course of an epidemic.
Nevertheless, several researchers~(\cite{barthBarrSatoVesp1,satoVesp2,satoVesp3,boguSatoVesp1,eameKeel1}) have been successful modeling many of the dynamical aspects of network epidemics, particularly in the early stage where asymptotically correct formuli for disease incidence are now known. We improve upon these results by presenting a system of nonlinear differential equations which can be used to solve for the complete incidence curve, as well as other quantities of interest. We treat the simplest possible case of the SIR dynamics with constant rate of infection and recovery. Section~\ref{sec:mode} describes the model. Several examples are given in section~\ref{sec:exam}.

\section{The model\label{sec:mode}}

	\begin{table}
	\begin{center}
		\begin{tabular}{l}
		\\
		\hline
		$ {\displaystyle  \dot{\beta } = \alpha~\mu~p_W }    $\\
		${\displaystyle \dot{\alpha} = -\alpha (r+\mu) p_W } $  \\
		${\displaystyle \dot{T} = -(r + \mu) p_W T -  p_W~r~n~\alpha^2 g''(\alpha + \beta) } $\\
		${\displaystyle \dot{W} = p_W  (  r~n~\alpha^2  g''(\alpha + \beta) - (r+\mu)(2W+T) )  }$ \\
		\hline
		\end{tabular}
		\caption{A summary of the nonlinear differential equations used to the describe the spread of a simple SIR type epidemic through a random network. The degree distribution of the network is generated by $g(x)$.}
	\end{center}
	\end{table}

We investigate undirected random networks with specified degree distributions\footnote{The \emph{degree} of a node in a network is the number of connections to that node. The \emph{degree distribution} is a discrete probability density over the positive integers describing the probability of realizing a given degree.}\cite{strog1}. Let $p_k$ be the probability of a node having a degree k. As in previous research we will make great use of the probability generating function (PGF) corresponding to the degree distribution. 

Although widely employed in the probability theory and the study of stochastic branching processes, generating functions are less familiar to those working in mathematical epidemiology (but see\cite{beck1,fa,al,an}). The utility of PGF's for the current investigation cannot be understated. Consider the degree distribution \emph{among susceptibles} at a given time t. As an epidemic progresses, more highly connected nodes, often called ``hubs'', will be preferentially culled from the population of susceptibles. Thus the degree distribution among susceptibles will evolve as the epidemic progresses. Our approach will be to keep track of the evolution of this distribution by careful application of parameters to the PGF. This will ultimately allow us to find the number of infecteds at any given time. 

Given a degree distribution, we define the probability generating function $g(x)$ as
\begin{equation}
	g(x) = p_0 + p_1 x + p_2 x^2 + p_3 x^3 + \cdots
\end{equation}
In most cases this series will converge to an algebraic function, in which case any operation to be done on the PGF can be done on the simple algebraic form. The series form can be retrieved by Taylor expansion. The degree distribution is a parameter of the model, so $g$ must be well-defined. Several examples for common distributions are given in section~\ref{sec:exam}. The results given below hold for any degree distribution. 

It will be helpful to the reader if several examples are provided to illustrate the utility of PGF's. Generating functions allow us to manipulate probability densities using simple algebraic operations. For example, if we were to draw two realizations of a random variable $X$ with generating function $g(x)$, the density of the sum would have generating function $\sum_k (p_1p_{k-1} + p_2p_{k-2}+\cdots)x^k = g^2(x)$. The mean of the random variable can be be computed by differentiating the generating function, $<X> = \sum_k k p_k  = g'(1)$. 

Another example more apropos to this paper is the following: Suppose we select a fraction $\alpha$ of the stubs\footnote{
In network vernacular a \emph{stub} is one end of a \emph{connection} between two nodes.
} from a network whose degree distribution has generating function $g(x)$. Then what proportion of nodes will \emph{not} be attached to any of the stubs we selected?
\begin{displaymath}
\sum_k p_k (1-\alpha)^{k}  = g(1-\alpha)
\end{displaymath}
Meanwhile the degree distribution of those not attached to a selected connection is generated by
\begin{displaymath}
{\displaystyle  \frac{g((1-\alpha) x )}{g(1-\alpha)}  }
\end{displaymath}
We can do better by computing the explicit generating function for the joint degree distribution of selected and unselected stubs. This is accomplished by applying a second variable to the generating function. Let $x$ correspond to selected stubs and $y$ correspond to unselected stubs. The probability of a degree $k$ node having $m$ of its stubs selected is $\binom{k}{m}\alpha^m(1-\alpha)^{k-m}$. Then the generating function will be of the form
\begin{eqnarray*}
	g(x,y;\alpha) = \sum_k \sum_{m=0}^k p_k\binom{k}{m}\alpha^m(1-\alpha)^{k-m} x^m y^{k-m} / c \\
	 = \sum_k p_k (\alpha x + (1-\alpha) y)^k / c = g(\alpha x + (1-\alpha) y) / c
\end{eqnarray*}
where $c = g(\alpha+\beta)$ is a normalizing constant. 
This example is important, as it underlies the methodology employed in this paper. The situation would be identical if  infection had spontaneously spread among a fraction $\alpha$ of the stubs and we asked how many nodes remained uninfected.

We will use an indirect approach in that we will not track the evolution of susceptibles and infecteds directly, but rather the number of stubs which are attached to susceptibles and infecteds. When an infected node transmits infection along one of its connections, we say the corresponding connection is \emph{occupied}. The variable $T$ will be the number of stubs emanating from susceptible nodes which are not paired with an infected or refractory alte. The variable $W$ will be the number of stubs emanating from infected nodes which have not yet become infected or refractory. We will treat the simple case of a constant force of infection and constant recovery rate. The quantities of interest in the model are as follows:%
\begin{itemize}
\item $r:=$  Force of infection. The probability per unit time of infection traversing a network connection.
\item $\mu:=$ Recovery rate. The probability per unit time that a connection to an infected will become refractory.
\item $n:=$ The population size.
\item $z:=$ The average degree in the network.
\item $T:=$  The number of all network connections to susceptible nodes which have not become refractory.
\item $W:=$  The number of all network connections to infected nodes which have neither become occupied nor become refractory.
\item $\alpha:=$ The proportion of stubs not connected to an occupied or refractory stub, i.e. the survivor function of susceptible stubs.
\item $\beta:=$ The proportion of stubs among susceptible nodes which are connected to refractory stubs.
\item $S:=$ The number of susceptibles.
\item $I:=$ The number of infecteds, including those in a refractory state.
\end{itemize}

It is important not to confuse stubs and connections. Two stubs are paired to form a connection. Stubs can be dormant, can be infected (infection has been transmitted by the stub to its alter), or can be refractory. In particular, it is possible for one stub to be refractory while its alter is infected. However if just one stub in a connection is infected, we say the corresponding connection is occupied. The dynamics proposed below do not keep track of the number of occupied connections, but rather of the number of stubs paired with infected or refractory alters. This is a pragmatic approach, as a susceptible can be defined as a node for which all of its stubs are not connected with infected alters. 

During the course of an epidemic, a node may be connected to a refractory stub, an infected stub, or a dormant stub. The different types of connections can be factored into the generating function by using multiple variables. Let the variable $x$ correspond to the number of stubs paired with dormant alters, and $y$ correspond to the number of stubs paired with refractory alters. Note that at any given time, a susceptible will not have any stubs connected to an infected alter by definition. Since we are only interested in the degree distribution of susceptibles, we will not introduce a variable for the number of infected stubs.

For susceptibles, stubs will be distributed among refractory connections and unoccupied/non-refractory connections. As defined above, $\alpha$ is the probability of having the latter type of connection, while $\beta$ is the probability of the former. The generating function for the degree distribution among susceptibles will be
\begin{equation}
\sum_k p_k (\alpha x + \beta y)^k / c = g(\alpha x + \beta y) /g(\alpha + \beta)  \label{eqn:dist1}
\end{equation}
The quantity $T$ is easy to derive by similar logic. The probability of a node having degree $k$ and contributing $m$ stubs to $T$ is
\begin{displaymath}
p_k\binom{k}{m}\alpha^m \beta^{k-m} 
\end{displaymath}
So in terms of the PGF, the number of stubs emanating from susceptibles which do not have refractory alters will be 
\begin{equation}
{\displaystyle T = n\frac{d}{d x} [ g(\alpha x + \beta y)]_{x =1, y=1}  = n~\alpha~g'(\alpha + \beta) } \label{eqn:T}
\end{equation}

$\alpha$ and $\beta$ will change over the course of the epidemic, thereby controlling the evolution of the degree distribution~(\ref{eqn:dist1}). It remains to determine the dynamics of these parameters.  At any given time, the hazard rate for an unoccupied stub being connected to an infected stub is $r p_W$, where $p_W = W/(W+T)$ is the proportion of non-refractory/unoccupied stubs connected to infecteds. Likewise, the hazard rate for becoming connected to a refractory stub is $\mu p_W$. Recall $\alpha$ is the survivor function for stubs not connected to occupied or refractory stubs; thus its dynamics is governed by
\begin{equation}
	{\displaystyle \dot{\alpha} = -\alpha (r+\mu) p_W } \label{eqn:alpha}
\end{equation}

The evolution of $\beta$ is more complicated. The probability of a stub connected to a susceptible node surviving to a time t is of course $\alpha$. At time t, the hazard of connecting to a refractory stub is $\mu p_W$. Then we have the following:
\begin{equation}
	{\displaystyle  \dot{\beta} = \alpha~\mu~p_W }  \label{eqn:beta}
\end{equation}

The dynamics of W is dependent both on the outflow of stubs becoming occupied and refractory, plus the inflow of stubs from newly infected nodes. Note that the total degree mass of the network, $M = n z$ is conserved. If we denote by $\mathcal{X}$ the stubs which are either occupied or refractory, we have the identity $W = M - T - \mathcal{X}$. Differentiating gives $\dot{W}$. 
\begin{equation}
\dot{W} = -\dot{T} - \dot{\mathcal{X}} \label{eqn:dw1}
\end{equation}
$\dot{\mathcal{X}}$ is quite simple. When a network connection becomes occupied or refractory, the two stubs making up the connection change state. Then $\dot{\mathcal{X}}$ increases at twice the rate at which stubs from W become refractory or occupied.
\begin{displaymath}
\dot{\mathcal{X}} = 2(r+\mu) W \label{eqn:dx}
\end{displaymath}
Differentiating equation~(\ref{eqn:T}) and using equation~(\ref{eqn:alpha}) gives
\begin{equation}
{\displaystyle \dot{T} = -(r + \mu) p_W T -  p_W~r~n~\alpha^2~g''(\alpha + \beta) } \label{eqn:dT}
\end{equation}
Finally, combining equations~(\ref{eqn:dw1}),~(\ref{eqn:dx}), and~(\ref{eqn:dT}) we have
\begin{eqnarray}
{\displaystyle \dot{W} = (r+\mu)(p_W T - 2W) +  p_W~r~n~\alpha^2 g''(\alpha + \beta) } \\
{\displaystyle = p_W  (  r~n~\alpha^2  g''(\alpha + \beta) - (r+mu)(2W+T) )  } \label{eqn:dW}
\end{eqnarray}
This completes the model.

Once the model has been integrated the number of susceptibles can be determined by applying the PGF to distribution parameters $\alpha$ and $\beta$. At a given time t, the number of susceptibles S is
\begin{eqnarray}
	S = n \sum_k \sum_{m=0}^k p_k\binom{k}{m} \alpha^m \beta^{k-m}    \\
	 = n \sum_k p_k (\alpha  + \beta )^k  = n~g(\alpha + \beta) \label{eqn:S}
\end{eqnarray}
The number of infecteds including those who have recovered is $I = n - S$.

\section{Examples\label{sec:exam}}

The model has been tested on several common degree distributions:
        \begin{itemize}
                \item Poisson: $p_{k} = \frac{z^{k} e^{-z}}{k!}$. This is generated by
                \begin{equation}
                        g(x) = e^{z (x - 1)}  \label{eqn:poisson}
                \end{equation}
                \item Power-law. For our experiments, we utilize power-laws with exponential cutoffs $\kappa$: $p_{k} = \frac{k^{-\gamma}e^{-k/\kappa}}{Li_{\gamma}(e^{-1/\kappa})}, k\geq 1$ where $Li_{n}(x)$ is the nth polylogarithm of x. This is generated by
		\begin{equation}
			g(x) = Li_\gamma(x e^{-1/\kappa}) / Li_\gamma(e^{-1/\kappa})  \label{eqn:powerlaw}
		\end{equation}
                \item Exponential: $p_{k} = (1-e^{-1/\lambda}) e^{-\lambda k}$. This is generated by
                \begin{equation}
                        g(x) = \frac{1 - e^{-1/\lambda}}{1 - x e^{-1/\lambda}} \label{eqn:exponential}
                \end{equation}
        \end{itemize}
	
\begin{figure}
		\begin{center}
			\includegraphics[width = .75\textwidth]{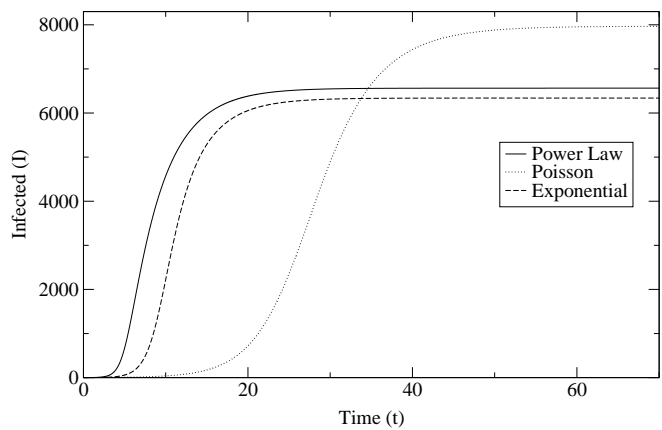}
			\caption{ The number of infecteds (including recovered) is shown versus time for an SIR model on three networks. Force of infection and mortality are constant: $r=0.2$, $\mu = 0.1$. The networks have Poisson ($z = 3$), power law ($\gamma = 1.615, \kappa = 20$), and exponential ($\lambda=3.475$) degree distributions. Each of these degree distributions has an average degree of 3.  }
			\label{fig:incidence}
		\end{center}
	\end{figure}

\begin{figure}
		\begin{center}
			\includegraphics[width = .5\textwidth]{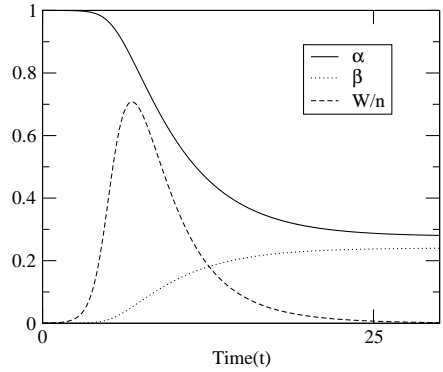}
			\caption{$\alpha$, $\beta$, and $W/n$ are shown versus t for a power law network with exponent $\kappa = 1.615$ and exponential cutoff $\kappa = 20$. Force of infection and mortality are constant: $r=0.2$, $\mu = 0.1$. }
			\label{fig:misc}
		\end{center}
	\end{figure}

If a single node is chosen at random from the population and infected, we can anticipate the following initial conditions: The survivor function for uninfected stubs, $\alpha$,  will begin at 1 and evolve downwards. $\beta$ will begin at 0 and evolve upwards. T will be equivalent to the degree mass of the network minus the degree of the initial infected. And W will be the degree of the initial infected. We take the degree of the initial infected to be the average degree within the network. These are the initial conditions used in the trials shown in figure~\ref{fig:incidence} and~\ref{fig:misc}. 

Figure~\ref{fig:incidence} shows the disease incidence for each of the degree distributions~(\ref{eqn:poisson}),~(\ref{eqn:powerlaw}), and~(\ref{eqn:exponential}), with a force of infection $r=.2$ and mortality $\mu=.1$. The parameters of the degree distributions were chosen so that each network has an identical average degree of 3.  That is, the density of connections in each network is the same. Nevertheless, there is widely different epidemic behavior due to the different degree distributions. 

A sense for the different dynamical behaviors of each of the three networks can be gotten from figure~\ref{fig:incidence}. Consistent with previous research, the degree distribution has a great impact on the final size of the epidemic~(\cite{meyePourNewmSkowBrun1,newm1}). More importantly, the three networks exhibit widely varying dynamical behavior.
In particular, note that the power law network experiences epidemics which accelerate very rapidly. Such epidemics enter the expansion phase virtually as soon as the first individual in the network is infected. Both the Poisson and exponential networks experience a lag before the expansion phase of the epidemic. These observations are consistent with the findings in~(\cite{barthBarrSatoVesp1}) that the timescale of epidemics shortens with increasing contact heterogeneity. Pure power laws have an infinite second moment, and therefore have a minimally short time-scale. This has important implications for intervention strategies, as it is often the case that interventions are planned and implemented only after a pathogen has circulated in the population for some time. If an epidemic were to occur in the power-law network, there would be little time to react before the the infection had reached a large proportion of the population.

Several other variables of interest are computed as a byproduct of the model. Figure~\ref{fig:misc} shows the most important for the power law trial described above. $\alpha$ shows the proportion of stubs not connected to an occupied or refractory alter. $\beta$ shows the proportion of stubs \emph{among susceptibles} connected to a refractory alter. These variables do not quite move in tandem and may cross each other. Also shown is $W$ (rescaled by population size $n$) which is similar to the hazard rate of becoming infected ($rW/(W+T)$). The epidemic ceases only when $W$ reaches negligible levels. 
	
\begin{figure}
		\begin{center}
			\includegraphics[width = .5\textwidth]{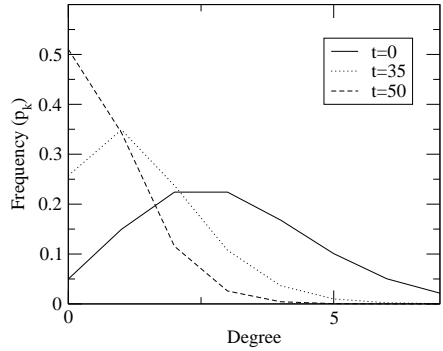}
			\caption{ The degree distribution (equation~(\ref{eqn:dist2})) for susceptibles is shown at three different times during the course of an epidemic on a Poisson network ($z=3$). Force of infection and mortality are constant: $r=0.2$, $\mu = 0.1$. }
			\label{fig:dist}
		\end{center}
	\end{figure}

Something offered by this model and not to the author's knowledge seen previously, is an explicit calculation for how the degree distribution of susceptibles evolves over the course of the epidemic. The infection will clearly tend to strike more highly connected individuals before more isolated individuals. Thus we expect the degree distribution to become bottom-heavy, as high degree nodes are gradually weeded out of the population. This is indeed observed in figure~\ref{fig:dist} for the Poisson trial described above. 

Recall that the degree distribution of susceptibles is generated by equation~(\ref{eqn:dist1}) and that we retrieve the explicit degree distribution by differentiation:
\begin{equation}
{\displaystyle p_k = [ (\frac{d^k}{dx^k} g(x) ]_{x=0}  / k!} 
\end{equation}
Applying this to the Poisson PGF (equation~(\ref{eqn:poisson})) gives
\begin{equation}
{\displaystyle   p_k = \frac{(z\alpha)^ke^{-z\alpha}}{k!}  } \label{eqn:dist2}
\end{equation}
We recognize this as simply the Poisson distribution with an adjusted parameter $z\times\alpha$.   
	
\begin{figure}
		\begin{center}
			\includegraphics[width = .5\textwidth]{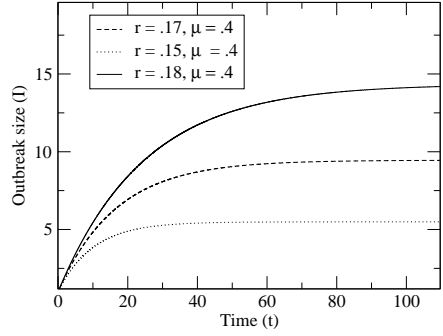}
			\caption{ The number of infecteds (including recovered) is shown versus time for an SIR model on a Poisson network ($z=3$). Each of these trials are below the critical level of transmissibility required to sustain an epidemic. Mortality is constant, $\mu = 0.4$, while three different levels of the force of infection are tried, $r=0.15,0.17,0.18$.}
			\label{fig:outbreak}
		\end{center}
	\end{figure}

Previous work~(\cite{newm1}) has shown that there is a critical transmissibility above which an epidemic will reach a fraction of the population in the limit as $n$ goes to infinity. Below that threshold, the epidemic is limited to a finite-sized outbreak. Figure~\ref{fig:outbreak} shows the qualitatively different dynamical behavior of outbreaks below the phase transition for networks with a Poisson distribution. Note that these outbreak sizes are independent of the population size, $n$, in contrast to the incidence curves above the phase transition which are sensitive to $n$ both in the time-scale of the epidemic and the number ultimately infected. 

Define the \emph{transmissibility}, $\tau$, of the disease as the probability that the infection will traverse a network connection between and infected and a susceptible\footnote{$\tau$ is related to to the traditional $R_{0}$ through the degree distribution. See~\cite{meyePourNewmSkowBrun1}}. With constant force of infection and mortality
\begin{displaymath}
	{\displaystyle  \tau = \frac{r}{r + \mu}  }
\end{displaymath}
What is the critical transmissibility that defines the phase transition? Recall that the epidemic is complete when $W$ is negligible and decreasing. If $W$ is decreasing at $t=0$ then the epidemic will necessarily die out without reaching a fraction of the population. The critical point occurs where
\begin{displaymath}
\dot{W}_{t=0} = 0 = -\dot{T} - \dot{\mathcal{X}}
\end{displaymath}
Applying equations~(\ref{eqn:dT}) and~(\ref{eqn:dx})
\begin{eqnarray*}
{\displaystyle  0 = \frac{\alpha W}{W + T}[(r+\mu)g'(\alpha + \beta) + \alpha~r~g''(\alpha+\beta)] - 2(r + \mu)   }  \\
{\displaystyle  \frac{r+\mu}{r}\Big( \frac{\alpha n}{W+T}g'(\alpha + \beta) - 2 \Big) =  \frac{-\alpha^2~n~g''(\alpha+\beta)}{W+T}  } \\
{\displaystyle    \frac{r}{r+\mu} = \tau = \frac{2W+T}{n~\alpha^2g''(\alpha+\beta)}   }
\end{eqnarray*}
At $t=0$, $\alpha = 1$, $\beta = 0$, $W\approx 0$ and $T\approx n~g'(1)$. Then
\begin{equation}
	\tau^* = g'(1) / g''(1)
\end{equation}
This is in agreement with previous results based on bond-percolation theory~(\cite{newm1}). 

	
\section{Discussion\label{sec:disc}}

The statistical properties of SIR epidemics in random networks have been understood for some time, but the explicit dynamics have been understood mainly through simulation. This paper has addressed this shortcoming by proposing a system of differential equations to model SIR in random networks.  

It should be noted that the SI dynamics are a special case of this model ($\mu=0$), in which case the ultimate extent of the epidemic is simply the giant component~(\cite{mollReed1})\footnote{
The \emph{giant component} of a network, if it exists, is the largest set of nodes such there exists a path between any two of them; furthermore the giant component must occupy a fraction of the network in the limit as network size goes to infinity.}
 of the network.

The distribution of contacts, even holding the density of contacts constant, has enormous impact on epidemic behavior. This goes beyond merely the extent of the epidemic, but as shown here, the dynamical behavior of the epidemic. In particular, the distribution of contacts plays a key role in determining the onset of the expansion phase. 

The distribution dynamics from equation~(\ref{eqn:dist1}) and shown in figure~\ref{fig:dist} have important implications for vaccination strategies. Previous work~(\cite{kaplCrafWein1,hallLongNizaYang1}) has focused on determining the critical levels of vaccination required to halt or prevent an epidemic. It is usually taken for granted that contact patterns among susceptibles are constant. Furthermore, most widespread vaccinations occur only once an epidemic is underway. Future research could be enhanced by considering optimal vaccination levels when the epidemic proceeds unhindered for variable amounts of time. 

It is hoped that the distribution dynamics described in this paper will find applications beyond modeling heterogeneous connectivity. The dynamic PGF approach may be used to capture other forms of heterogeneity, such as of susceptibility, mortality, and infectiousness.  


\end{document}